# Novel utilization of a paper-level classification system for the evaluation of journal impact: An update of the CAS Journal Ranking

**Sichao Tong[1,2], Fuyou Chen[1], Liying Yang[1], Zhesi Shen[1,*]**

1. National Science Library, Chinese Academy of Sciences, 100190, Beijing, China

2. Department of Information Resources Management, School of Economics and Management, University of Chinese Academy of Sciences, 100190, Beijing, China

Since its first release in 2004, the CAS Journal Ranking, a ranking system of journals based on a citation impact indicator, has been widely used both in selecting journals when submitting manuscripts and conducting research evaluation in China This paper introduces an upgraded version of the CAS Journal Ranking released in 2020 and the corresponding improvements. We will discuss the following improvements: (1) the CWTS paper-level classification system, a fine-grained classification system, utilized for field normalization, (2) the Field Normalized Citation Success Index (FNCSI), an indicator which is robust against not only extremely highly cited publications, but also wrongly assigned document types, and (3) document type difference. In addition, this paper will present part of the ranking results and an interpretation of the features of the FNCSI indicator.

[*] Corresponding author: Zhesi Shen(shenzhs@mail.las.ac.cn)

Sichao Tong(0000-0003-0658-9963), Fuyou Chen(0000-0001-5923-7086), Liying Yang(0000-0001-5539-9934), Zhesi Shen(0000-0001-8414-7912)

# 1. Introduction

CAS Journal Ranking[1], an annually released journal ranking by the Center of Scientometrics (CoS), National Science Library of Chinese Academy of Sciences (CAS), is a journal ranking system widely recognized in China. It ranks journals included in Clarivate's Journal Citation Reports (JCR), based on citation indicators. In this paper, we will sketch its history and introduce the upgraded version of the CAS Journal Ranking.

In Section 1.1 we will briefly describe the history of CAS Journal Ranking, then we will introduce the old version of CAS Journal Ranking in Sec.1.2. In Sec.1.3 we point out its limitations, and corresponding improvements. A literature review of normalized journal indicators is presented in Sec. 2. Data and method used in the upgraded version of CAS Journal Ranking are presented in Sec.3. Ranking results and characteristics of the upgraded version are described in Sec.4. In the Conclusion and Discussion section, we will discuss the usage of the CAS Journal Ranking and some future directions.

## 1.1 History of CAS Journal Ranking

The origin of the CAS Journal Ranking is related to the misuse of the journal impact factor (JIF) in China. In 1963, the idea of a JIF to measure journal impact was first proposed by Garfield and Sher (Garfield & Sher, 1963). Afterward, the JIF was gradually accepted in the sciences and some social sciences. In the late 1990s, the JIF was first introduced to China, and provided a reference for the Chinese research community, especially for junior researchers, to estimate the quality of journals indexed in the Web of Science (WoS). At that time, one feature of the JIF, namely that the JIF varies significantly across disciplines, was ignored by the Chinese research community and policymakers. So, what happened was that JIFs from different disciplines were compared. Beginning in 2000, in order to promote the proper use of JIF in terms of heterogeneity across different disciplines, the CoS started to share the CAS Journal Ranking with some Chinese universities. In 2004, the first edition of the CAS Journal Ranking (the old CAS Journal Ranking) was officially released and continually updated until now. After years of developing the ranking method and indicators, the upgraded CAS Journal Ranking was released in January 2020.

[1] The website (https://www.fenqubiao.com/) is for registered institutional users. For individual researchers, please scan the QR code on the website and follow the Wechat official account.

The original intention of CAS Journal Ranking was to provide a reference for researchers when selecting journals for publishing. Twenty years ago, the degree of internationalization in China was much lower than it is now. A considerable number of researchers, especially junior researchers, rarely had the opportunity to participate in international communication, so it was hard for them to interpret journal impact metrics (e.g., JIF). After the first release, the CAS Journal Ranking was approved and became widespread in China. Later in 2007, the JCR released its quartile ranking. A survey[2] among Chinese researchers conducted by Elsevier's STM Journals China Program team in 2021 shows that the CAS Journal Ranking is the most recommended journal list in China. Moreover, several CAS institutions and Chinese universities utilized the CAS Journal Ranking as one of the references to evaluate the citation impact at the institutional level.

How to utilize quantitative data properly in research performance evaluation is a challenge for many countries, not just for China. Although the original intention of the CAS Journal Ranking is not relevant to this task, the CAS Journal Ranking has, inevitably, been involved in this controversy. This is a problem that CAS Journal Ranking has to deal with. However, in this paper, we will not discuss research evaluation at large but focus on the improvement of the journal evaluation method from a methodological aspect.

## 1.2 The old CAS Journal Ranking

The main idea of the CAS Journal Ranking is to group journals in each discipline category into tiers, following the idea of Jin and Wang (1999). The primary method of the old CAS Journal Ranking is as follows. Firstly, WoS journals were grouped by discipline category. Secondly, the journals of each category are divided into four tiers based on the descending order of their indicators. Here we use the average of recent 3 JIFs to rank journals, i.e., $\mathrm{IF_Y} = (\mathrm{JIF_Y} + \mathrm{JIF_{Y\text{-}1}} + \mathrm{JIF_{Y\text{-}2}})/3$ . The top 5% of journals are classified as Tier 1, and then the rest journals are classified into Tier 2 to 4 to make sure that the total impact in each tier is equal. The distribution of impact indicators for each category varies somewhat, so the fraction of journals in each tier is slightly different. Yet, they still roughly have a pyramid-like structure (5%-20%-45%-100%). Journals within the same tier can be compared across disciplines. The idea of putting journals into tiers and comparing journals using tiers is adopted in the upgraded version of the CAS Journal Ranking.

2

https://www.elsevier.com/connect/behind-the-rising-influence-of-chinese-research?utm_source=twitter&utm_medium=social&utm_campaign=dt

## 1.3 Improvements in the upgraded CAS Journal Ranking

In the old CAS Journal Ranking, three significant limitations exist. One is related to JIF. For a journal, because the citation distributions are skewed (P. O. Seglen, 1992; P. O. Seglen, 1997; Lariviére, Kiermer, et al., 2016; Milojević, Radicchi, & Bar-Ilan, 2017), JIF can be vastly affected by the tail of highly-cited papers and may thus vary a lot across years. We previously utilized a three-year average JIF to alleviate such fluctuations. However, this solution is still not robust enough against occasionally highly-cited papers and cannot accurately reveal the average impact of journals.

The second limitation is that the citation potential varies greatly among different document types (Price, 1965). Research articles generally have a lower citation potential than review papers. Therefore, journals with a higher proportion of reviews can attract more citations. Hence, it is unfair to compare journals with different proportions of reviews. In the old version, we tried to solve this problem by ranking research journals and review journals separately. However, many research journals also publish a relatively large percentage of review articles.

The third limitation is that the discipline category is not fine-grained enough in the old CAS Journal Ranking. Initially, CAS Journal Ranking adopted a 13-field category, including Medicine, Physics, Biology, etc. Indeed, citation differs significantly within fields (see Fig.1). The JCR subject category, a more fine-grained category than 13 fields, was adopted by CAS Journal Ranking in 2008 to eliminate citation differences within these 13 fields. However, the problem still existed in the JCR subject category (van Eck, Waltman, van Raan, Klautz, & Peul, 2013) (see Fig.2 below). This limitation will reduce comparability for journals belonging to the same category.

To illustrate the effect of the third limitation, we plotted a scatter map of journals from all fields (see Fig.1), with each dot representing a journal and the color representing its potential citation.
For this map, we continue to use the same layout based on the journal citation network in the earlier work (Shen, Chen, Yang, & Wu, 2019). Moreover, we use the journal's expected JIF to indicate its potential citation. Regarding expected JIF, we also want to briefly introduce two basic definitions in Waltman's study (2016) as below.

- Expected number of citations for a publication: the average number of citations for total publications from the related topic, the same year, and the same document type; here, we use article or review type.
- Moreover, for a journal, the expected JIF is the average of expected number of citations, for the total publications of the journal.

Please refer to the ***Method and Data*** section for a more detailed formula for expected JIF. Figure 1 indicates a clear distinction in the potential citation among research fields. We can see that citations differ between different areas within the above-studied medical fields and many other fields; for example, the upper part and the lower part of the Math category obviously perform differently.

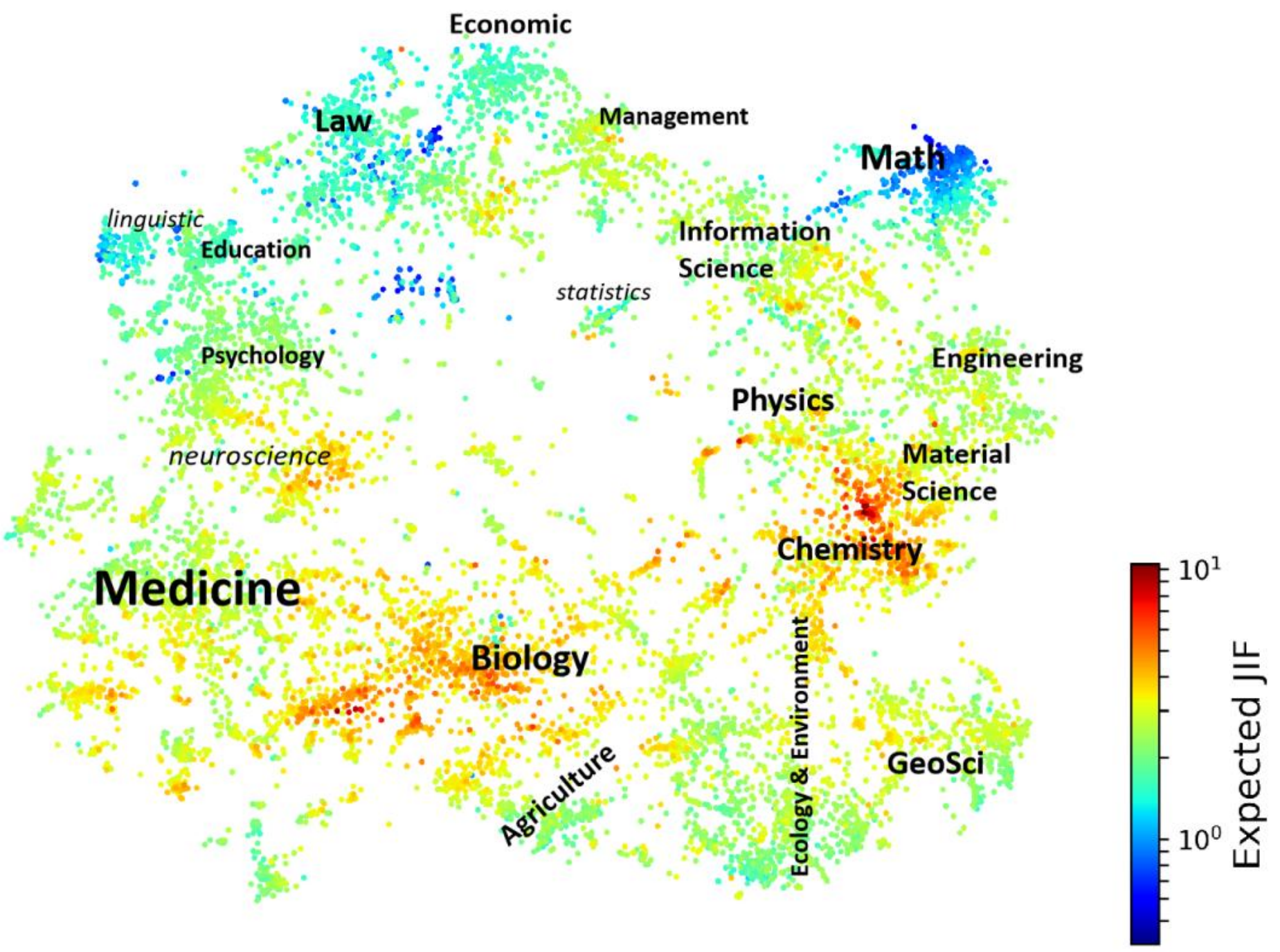

**Figure 1: Map of scientific journals with expected JIF.** *The color of each dot is related to the value of the corresponding journal's expected JIF: the darker the color of red/blue is, the larger/smaller the value of the expected JIF is.*

We then take journals from the JCR category-*Statistics & Probability* as an example. In Fig.2, each dot represents a journal, we color journals titled with *probability* in blue. We can find that most blue dots have smaller expected JIF, indicating a distinction of citation potential between journals from different topics within the *Statistics & Probability* category, e.g., probability-related journals have much smaller citation potential.

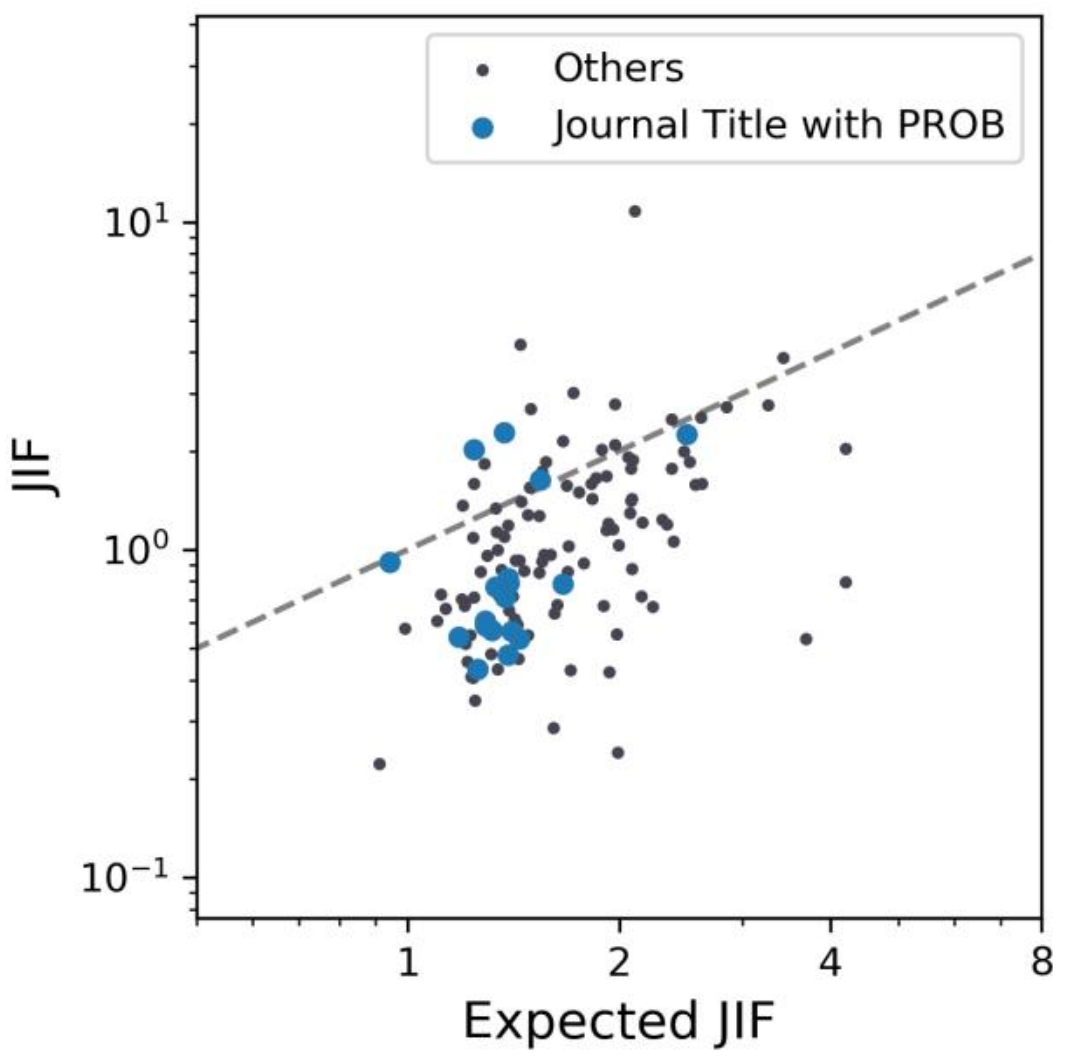

**Figure 2: Correlation of JIF and expected JIF for journals in the *Statistics and Probability* category.** *The dashed line is y=x, a dot positioned above the dashed line means that its actual JIF is larger than its expected JIF, and vice versa.*

To overcome the limitations mentioned above, we released the upgraded version of CAS Journal Ranking. This version includes the following refinements:

- Instead of JIF, a new indicator, Citation Success Index (CSI), was used. Compared with other citation indicators, for example, the three-year average JIF shown in earlier editions, CSI excels not only in robustness to ultra-small numbers of extremely highly cited publications but also in robustness to wrongly assigned document types.

- In addition, we consider the document type when performing normalization, i.e., calculating the indicator of articles and reviews, respectively.

- The CWTS paper-level classification system, a more fine-grained system, has been utilized to classify each paper into the corresponding cluster (topic) (see ***Method and Data*** section), to ensure the CSI indicator's calculation from a paper level.

## 2. Literature Review

In this section, we will briefly review the development of normalized journal indicators and embed CAS Journal Ranking in this developing timeline. The literature review is mainly based on the review by Waltman (2016). The normalization of journal indicators is mainly towards disciplinary differences, document types, and the skewness of citation distribution. In Fig.3, we present several representative works on normalized journal indicators related to CAS Journal Ranking.

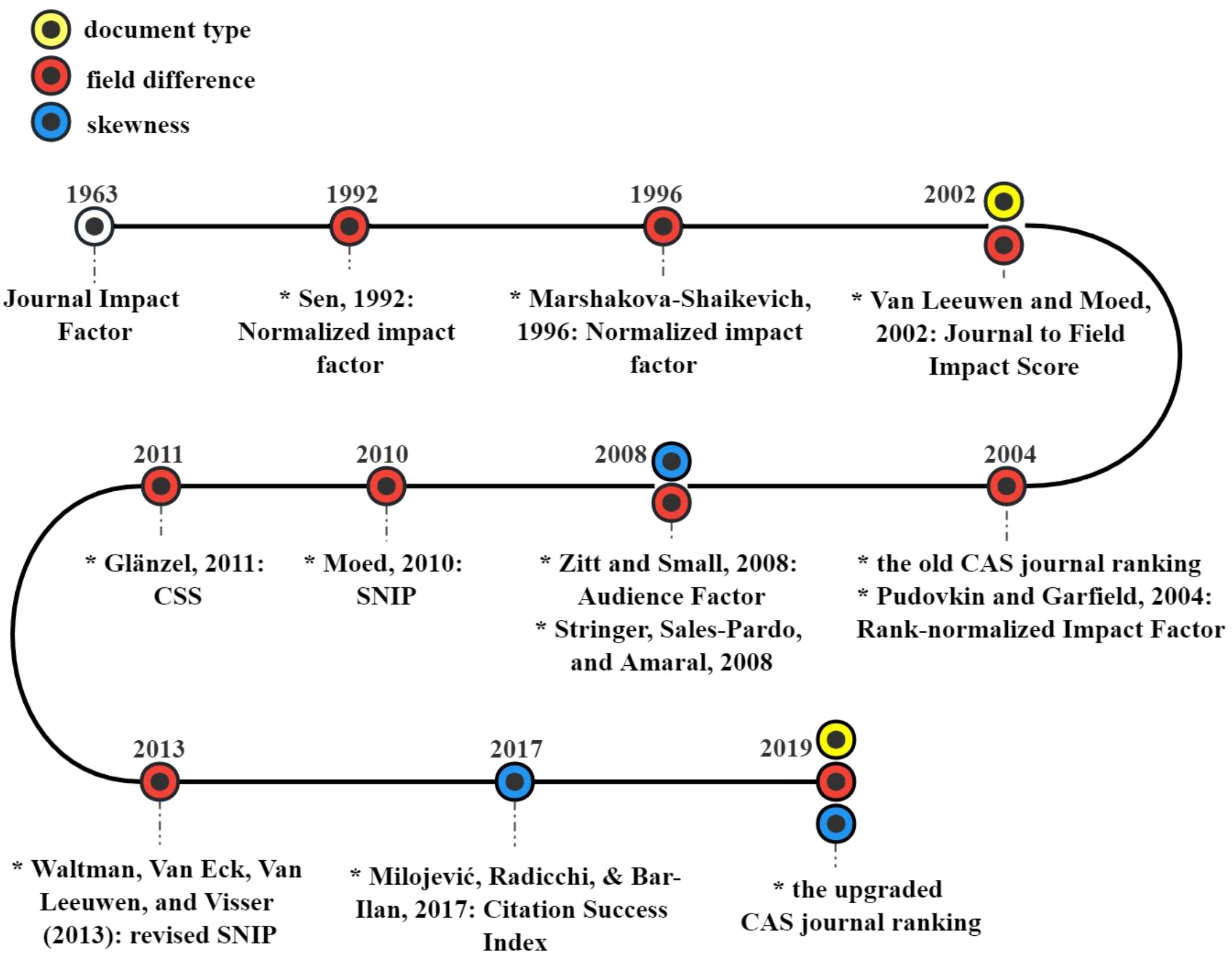

**Figure 3: Timeline of normalized journal indicators with CAS Journal Ranking embedded.**

Considering the field difference of JIFs, Sen (1992) and Marshakova-Shaikevich (1996) proposed normalizing JIF using maximum JIF or a few of the highest JIFs in each subject. Van Leeuwen and Moed (2002) proposed Journal to Field Impact Score considering the normalization of the document type, field, and citation window. Jin and Wang (1999) proposed putting journals in each category into 3 tiers of equal size and comparing journals across disciplines using tiers. In 2004, Pudovkin and Garfield (2004) proposed the rank-normalized impact factor (rnIF), which uses journals' relative position ordered by JIF within each JCR category to facilitate comparison among fields. In the same year, the first version of CAS Journal Ranking was released and proposed to group journals within categories based on JIF into four tiers and compare journals using tiers across fields. In 2011, Glänzel (2011) proposed normalizing JIF using the parameters extracted from Characteristic Scores and Scales (CSS). Zitt and Small (2008) proposed the audience factor, which uses the journal-level citing-side (or source-side) normalization. Later, Moed (2010) presented the Source Normalized Impact per Paper (SNIP), and Waltman et al. (2013) introduced a revised SNIP indicator. Toward the problem of skewness, researchers turn to the ranking-based indicators, e.g., percentile rank (Pudovkin & Garfield, 2009), percentile rank classes (Leydesdorff, Bornmann, Mutz, & Opthof, 2011; Bornmann,

Leydesdorff, & Mutz, 2013), the proportion of highly cited papers, indicators added in the Leiden Ranking (Waltman et al., 2012) and Clarivate's InCites[3], in which relative ranking matters instead of its absolute citation value. Stringer, Sales-Pardo, and Amaral (2008) proposed the probability that a randomly selected paper published in one journal has received more citations than a randomly selected paper published in another journal to test the effectiveness of comparing journals using JIF. Later Milojevic et al. (2017) defined this probability as Citation Success Index (CSI) and found an S-type relation between JIF and CSI. Shen et al. (2018) found that the relationship between JIF and CSI mainly results from the lognormal distribution of citations. Another way to alleviate the skewness problem is logarithmic conversion (Lundberg, 2007).

In the past, most colleagues used a journal-level classification system (e.g., WoS Subject Category) for normalization. However, the drawbacks of these classification systems for normalization have been revealed on many occasions. With the improvement of the accessibility of large-scale bibliometric datasets and an increase in computing power, data-based paper-level classification is constructed and used for normalization gradually (Waltman & van Eck, 2012, 2013a, 2013b). Towards the problems of journal indicators and combining the recent advances in scientometric methods, CAS Journal Ranking released its upgraded version in 2020.

# 3. Method and Data

This paper will take the 2019 version of the CAS Journal Ranking as an example to show the method and data, as well as the results.

## 3.1 Journals and citation data

CAS Journal Ranking (2019 version) includes the journals contained in Clarivate’s Journal Citation Reports (JCR) (2019 version according to Clarivate[4]). For citation data, we use Journal Impact Factor contributing items released by the JCR. It contains citations in year Y of each article and review, published in years Y-1 and Y-2, which counted towards the journal’s impact factor.

## 3.2 Paper-level classification data

We utilized the results of CWTS paper-level classification, which means each paper belongs to a certain cluster (topic). The article and review papers in Web of Science

[3] https://incites.clarivate.com/

[4] https://jcr.clarivate.com/

Core Collection - Science Citation Index Expanded and Social Sciences Citation Index between 2000 and 2018 were collected. For the details of constructing the CWTS paper-level classification system, we refer to (Waltman & van Eck, 2012, 2013a, 2013b) for an exhaustive introduction to the classification methods for exploring the relatedness of publications and clustering them into groups. Depending on the granularity, this classification system consists of three levels - macro, meso, and micro levels. Here we use the micro-level with about 4,000 clusters.

It should be noted that in the released results for the CWTS paper-level classification system, trade journals, and some local journals are excluded, i.e., because their citation links are too weak. Since we try to include as many journals as possible, we retrieved the related records from the WoS for these excluded journals and put them back into the corresponding clusters based on the clusters of the retrieved related records using the majority rule. In total, the upgraded version includes 99% of articles and reviews indexed by JCR. From the perspective of journals, 98% of journals with more than 90% of their total publications are included.

## 3.3 Journal Ranking Indicators/journal indicators utilized in this article

In the upgraded version, we follow the idea of the Citation Success Index (CSI) and extend CSI to a field-normalized indicator. The original CSI presented to compare the citation capacity between two journals (Stringer et al., 2008; Milojevic et al., 2017; Shen et al., 2018), is defined as the probability of a randomly selected paper from one journal having more citations than a randomly selected paper from the other journal. Following the same idea, we propose the Field Normalized Citation Success Index (FNCSI). The FNCSI can be defined as the probability that the citation of a paper from a journal is larger than a random paper on the same topic and with the same document type from other journals. More details will be introduced in the section below.

Before the upgraded version was completed, in order to investigate the differentiation between the new indicator (FNCSI) and the old indicator (JIF), we proposed the Field Normalized Impact Factor (FNIF). It should be noted that we did not use FNIF in the upgraded version; it is only used here for comparison.

- **Field Normalized Citation Success Index (FNCSI)**

For journal $A$, the probability $P(c_a > c_o \mid a \in A, o \in O)$ that the citation of a paper from journal $A$ is larger than a random paper on the same topic and with the same document type from other journals is defined as below:

$$S_A = P(c_a > c_o | a \in A, o \in O) = \sum_{t,d} P\,(A^{t,d}) P(c_a > c_o | a \in A^{t,d}, o \in O^{t,d})$$

Where $c_a$ is the citation count of one paper from journal $A$, $t \in \{\text{topic}_1, \text{topic}_2, \ldots\}$, $d \in \{\text{article}, \text{review}\}$, $A^{t,d}$ represents the publications belonging to journal A in topic $t$ with document type $d$, and $O$ represents the publications from the other journals. For a specific research topic t, its FNCSI is calculated as below:

$$S_A^t = \frac{1}{N_{A^t}} \sum_d N_{A^{t,d}} \left[ \frac{\sum_{a \in A^{t,d}, o \in O^{t,d}} 1\,(c_a > c_o) + \sum_{a \in A^{t,d}, o \in O^{t,d}} 0.5\,(c_a = c_o)}{N_{A^{t,d}} N_{O^{t,d}}} \right]$$

where weight 1 is for $c_a > c_o$ and weight 0.5 for ties, i.e., $c_a = c_o$. Journal $A$ usually involves several research topics at the micro level of the CWTS paper-level classification system, then the total FNCSI of Journal A can be summed from its affiliated topics as below:

$$S_A = \frac{1}{N_A} \sum_t N_{A^t} S_A^t$$

where $N_{A^t}$ is the number of publications of journal $A$ in topic $t$. The FNCSI is based on a two-year citation window (but this window can be adapted).

For a better understanding of FNCSI, examples of FNCSI formulation represented in an analogy way are presented in Appendix A.

- **Field Normalized Impact Factor (FNIF)**

The Field Normalized Impact Factor (FNIF) uses the same classification system as FNCSI but uses the average citations based on a normalization approach, i.e., each citation is normalized by the average citation of papers in the same topic cluster and with the same document type. For instance, the FNIF of journal $A$ is defined as:

$$F_A = \frac{\sum_{t,d} \sum_{a \in A^{t,d}} c_a / \mu_{t,d}}{N_A}$$

where $\mu_{t,d}$ is the average citation of papers in topic $t$ with document type $d$. By comparing the results of FNCSI and FNIF, we can see the advantages of CSI.

- **Expected JIF**

As we mentioned earlier, for each journal, we use the expected JIF as an indicator of citation potential:

$$E_A = \frac{\sum_{t,d} \mu_{t,d} N_{A^{t,d}}}{N_A}$$

where $\mu_{t,d}$ is the average citation of papers in topic $t$ with document type $d$.

# 4. Results

## 4.1. Ranking Results

This section presents the results of the CAS Journal Ranking based on FNCSI and the comparisons with other indicators[5]. Table 1 shows the top 20 journals ranked according to FNCSI. Here we only list journals mainly publishing research articles. This list is dominated by Nature-titled journals, Lancet-titled journals, and Cell-titled journals. The top five journals are well-acknowledged in the natural and life sciences. The other journals belong to different fields and do not concentrate on a single field or narrow group of fields.

The corresponding rankings based on FNIF values of these top 20 journals are also presented in Table 1. Among these journals, the rankings of Cancer Cell, Nature Neuroscience, Cell Metabolism, and Nature Immunology are boosted most by the FNCSI indicator; they all climb more than 20 positions. Only Lancet Oncology shows a slight drop in position from the FNCSI indicator. Overall, journals from medical-related categories have a relatively big gap between these indicators. In Appendix Table C1, we present the top 20 journals according to FNCSI and FNIF, respectively.

The correlation among relative rankings by FNCSI and FNIF is shown in Fig.4, with values closer to 0 representing better-ranked journals. We can see that FNCSI and FNIF are highly correlated (spearman correlation: 0.98, p-value: 0.00). In the lower part of Fig.3, we highlight several journals that have worse rankings in FNCSI compared with FNIF. These journals share a common property in that they each have one or several highly cited papers, and a majority of poorly cited papers, e.g., Chinese Phys C has one paper cited more than 2000 times, but about 70% of papers are not cited[6]. This result is consistent with the difference in definition between FNCSI and FNIF.

[5] All journals' FNCSI results for 2019 version are available at 10.57760/sciencedb.08419.

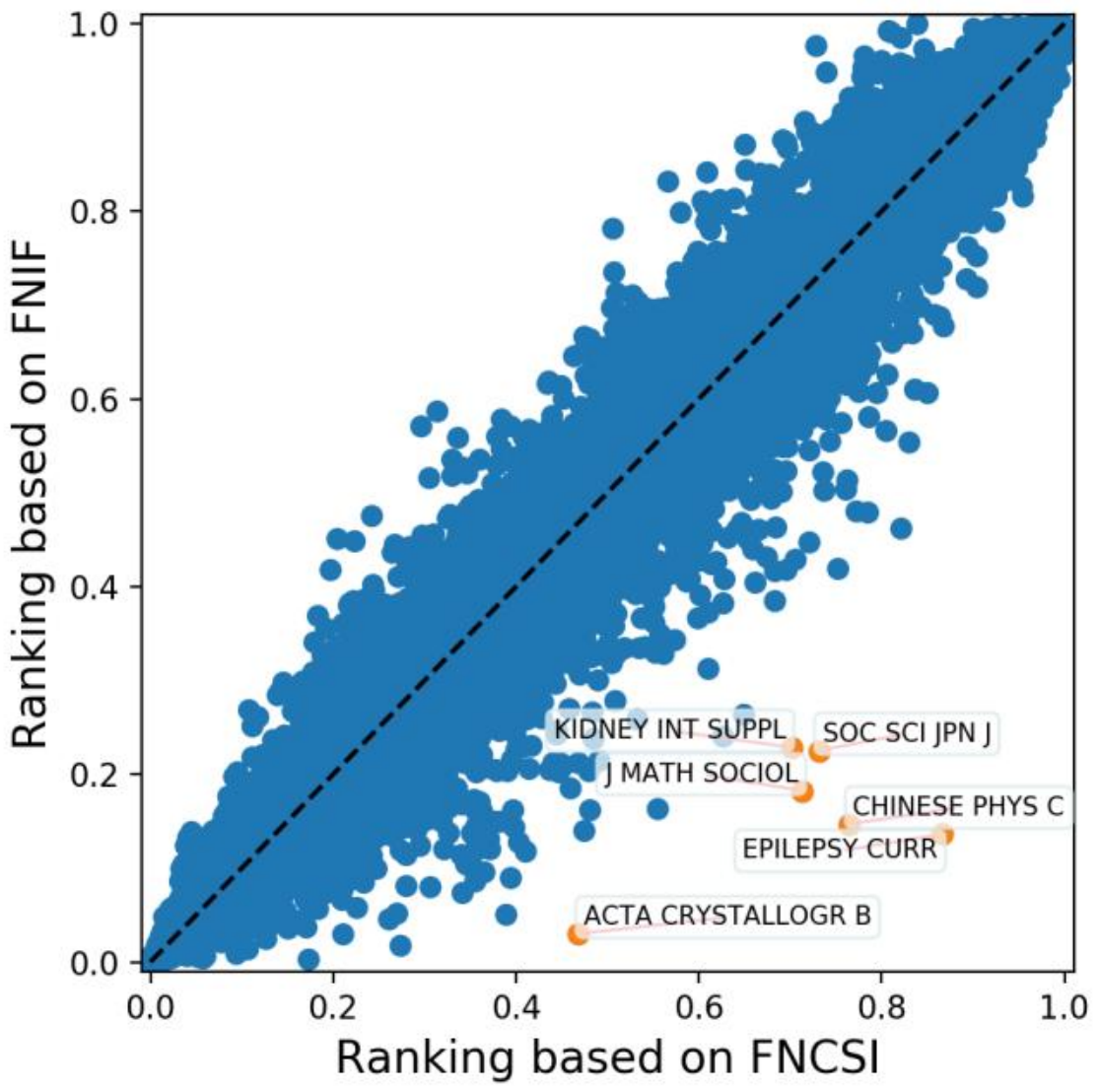

**Figure 4: Correlation of relative rankings based on FNCSI and FNIF.** *The smaller the better.*

**Table 1: Top 20 ranked journals according to FNCSI.**

| Journal | Category-WoS | FNCSI | FNIF |
|---|---|---|---|
| LANCET | MEDICINE, GENERAL & INTERNAL | 1 | 3 |
| NATURE | MULTIDISCIPLINARY SCIENCES | 2 | 5 |
| JAMA | MEDICINE, GENERAL & INTERNAL | 3 | 4 |
| SCIENCE | MULTIDISCIPLINARY SCIENCES | 4 | 9 |
| CELL | BIOCHEMISTRY & MOLECULAR BIOLOGY/CELL BIOLOGY | 5 | 15 |
| WORLD PSYCHIATRY | PSYCHIATRY | 6 | 8 |
| LANCET NEUROL | CLINICAL NEUROLOGY | 7 | 11 |
| NAT PHOTONICS | OPTICS/PHYSICS, APPLIED | 8 | 17 |
| NAT GENET | GENETICS & HEREDITY | 9 | 13 |
| NAT MED | BIOCHEMISTRY & MOLECULAR BIOLOGY/CELL BIOLOGY/MEDICINE, | 10 | 21 |

[6] https://jcr.clarivate.com/

| | | | |
|---|---|---|---|
| | RESEARCH & EXPERIMENTAL | | |
| NAT MATER | MATERIALS SCIENCE, MULTIDISCIPLINARY/CHEMISTRY, PHYSICAL/PHYSICS, APPLIED/PHYSICS, CONDENSED MATTER | 11 | 12 |
| LANCET ONCOL | ONCOLOGY | 12 | 10 |
| CANCER CELL | ONCOLOGY/CELL BIOLOGY | 13 | 38 |
| NAT CHEM | CHEMISTRY, MULTIDISCIPLINARY | 14 | 31 |
| NAT NEUROSCI | NEUROSCIENCES | 15 | 36 |
| CELL METAB | CELL BIOLOGY/ENDOCRINOLOGY & METABOLISM | 16 | 51 |
| LANCET RESP MED | CRITICAL CARE MEDICINE/RESPIRATORY SYSTEM | 17 | 22 |
| NAT IMMUNOL | IMMUNOLOGY | 18 | 58 |
| LANCET DIABETES ENDO | ENDOCRINOLOGY & METABOLISM | 19 | 27 |
| NAT NANOTECHNOL | NANOSCIENCE & NANOTECHNOLOGY/MATERIALS SCIENCE, MULTIDISCIPLINARY | 20 | 23 |

In section 1.3, we discussed the difference in citation potential among journals within the Statistics & Probability category, as shown in Fig.2. The paper-level classification system eliminates the effect of citation potential on journal rankings. In Table 2, we present the top 20 journals (mainly publishing research articles) according to FNCSI in this category. We can see that journals perform weakly in JIF due to low citation potential (low rank based on Expected JIF) revealed reasonably by FNCSI, such as several well-acknowledged journals like *Annals of Statistics*, *Annals of Probability*, and *Biometrika*.

**Table 2: Top 20 ranked journals in the Statistics and Probability category according to FNCSI. Their rankings based on Expected JIF and JIF are also presented.**

| Journal | Rank-FNCSI | Rank-Expected JIF | Rank-JIF |
|---|---|---|---|
| ECONOMETRICA | 1 | 69 | 2 |
| J R STAT SOC B | 2 | 45 | 3 |
| ANN STAT | 3 | 63 | 7 |
| PROBAB THEORY | 4 | 86 | 10 |
| REL | | | |
| ANN PROBAB | 5 | 103 | 15 |
| FINANC STOCH | 6 | 99 | 22 |
| J AM STAT ASSOC | 7 | 32 | 4 |
| INT STAT REV | 8 | 39 | 16 |
| J QUAL TECHNOL | 9 | 104 | 29 |
| J STAT SOFTW | 10 | 20 | 1 |
| ANN APPL PROBAB | 11 | 60 | 28 |
| STOCH ENV RES RISK A | 12 | 7 | 8 |
| BRIT J MATH STAT PSY | 13 | 9 | 20 |
| TECHNOMETRICS | 14 | 56 | 21 |
| BIOMETRIKA | 15 | 49 | 33 |
| BAYESIAN ANAL | 16 | 44 | 35 |
| BERNOULLI | 17 | 92 | 41 |
| INSUR MATH ECON | 18 | 65 | 43 |
| EXTREMES | 19 | 58 | 25 |
| ECONOMET THEOR | 20 | 91 | 52 |

## 4.2 Robustness

### 4.2.1 Robust against extremely highly cited publications

The robustness of an indicator represents its sensitivity to changes in the set of publications based on which it is calculated. A robust indicator will not change much against the occasional ultra-small number of highly cited publications. To measure the robustness of an indicator, we construct several sets of publications for each journal with the bootstrapping method and re-calculate the indicators and rankings accordingly. For instance, for a journal with N publications, we randomly select N publications with replacement, calculate these indicators, and get a new ranking for each journal. We simulate this procedure 100 times and thus obtain 100 rankings for each journal. Figure 5(a) shows the distribution of the obtained rankings of *Chinese Physics: C*. We can see that the ranking range of FNCSI is much smaller than FNIF. The citation distribution of *Chinese Physics: C* is highly skewed, with one paper cited about two thousand times and about 70% of papers not cited. Thus, FNIF depends strongly on whether this highly cited paper is included in the calculation or not.

To get an overview of the indicators' robustness, we calculate the relative ranking change for these indicators. The relative change in ranking is defined as:

$$\Delta = \frac{1}{N}\sum_{j}^{N}\frac{\max\{R_j\} - \min\{R_j\}}{\text{avg}\{R_j\}}$$

where $\{R_j\}$ is the rankings of journal j obtained from the above simulation. As shown in Fig.5(b), the relative change of FNCSI is smaller than FNIF implying that FNCSI is more robust than FNIF. FNCSI mainly focuses on the central tendency of the citation distribution and is not easily affected by occasional highly cited papers, which implies FNCSI can reflect the journal impact more stably.

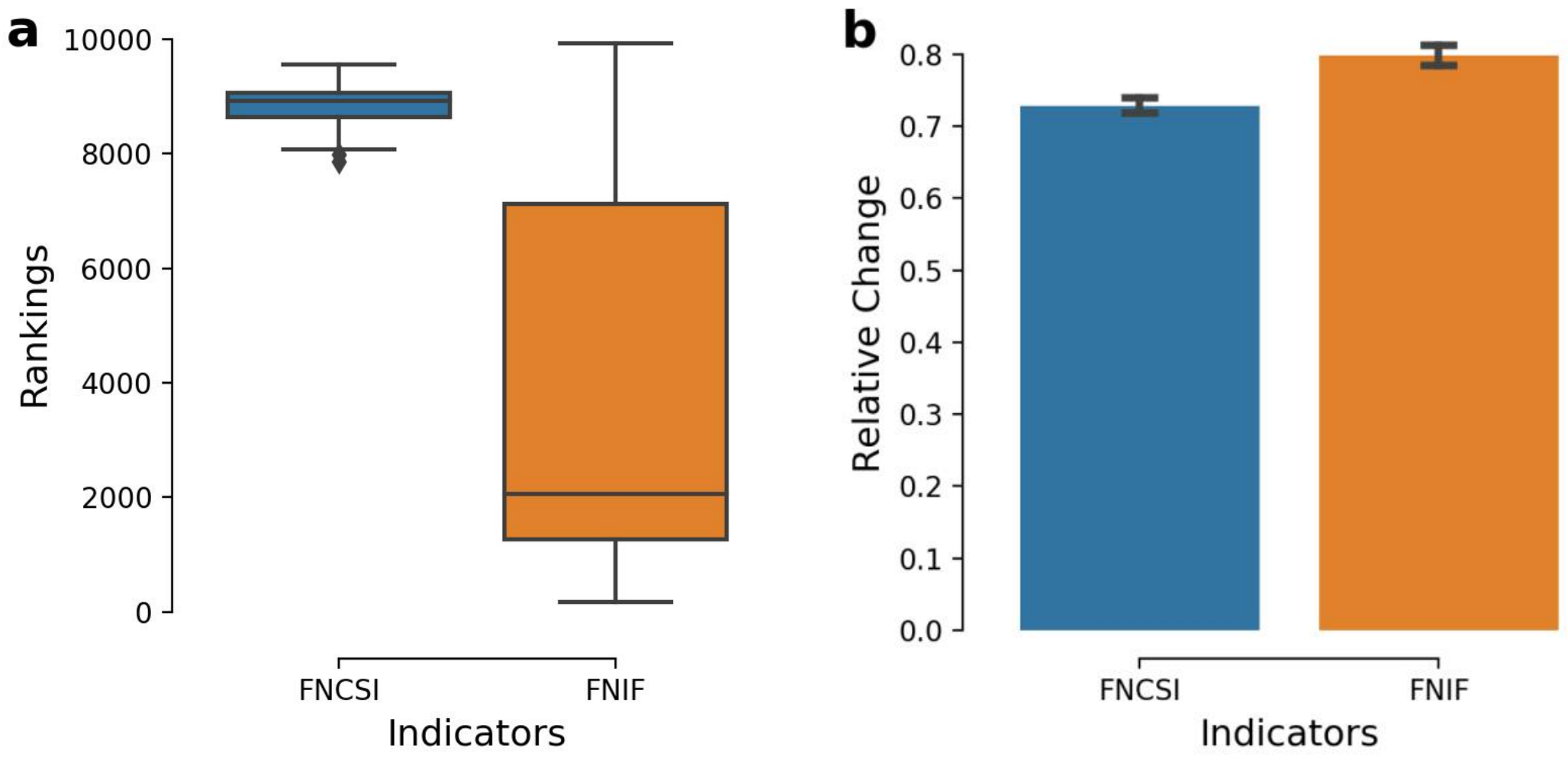

**Figure 5: (a) Ranking variability of *Chinese Physics: C* for FNCSI and FNIF. (b) average Relative change of rankings based on FNCSI and FNIF**

## 4.2.2 Robust against wrongly labeled Document Type

When conducting normalization for article/review, we need to assign the proper document type for each paper. Previous studies have shown that the document types assigned by WoS (here for articles and reviews) are not fully correct, e.g., some review papers are assigned as articles, and some articles are assigned as reviews (Colebunders & Rousseau, 2013; Harzing, 2013; Donner, P., 2017; Yeung, A. W. K., 2019; Zhu, Shen, Chen, & Yang, 2022). To test the sensitivity of indicators against wrongly labeled document types, here we generate a virtual data set:

- *Theoretically, we can randomly select N publications with replacement: article to Review or Review to Article. In practice, extreme cases were chosen to reveal this* *error's impact better. For each journal, we turn the document type of its most highly cited paper to the opposite, i.e., Article to Review or Review to Article.*

And then, we recalculate the journal indicators and obtain the new rankings based on FNCSI and FNIF, respectively. The comparison of rankings based on this changed data with the original rankings is shown in Fig.6. We can see that almost all the orange dots (FNCSI-based) locate closely along the diagonal line while the blue squares (FNIF-based) spread much broader, which implies that rankings based on FNCSI are more robust against wrongly labeled document type than rankings based on FNIF. From the point of fault tolerance, FNCSI performs better than FNIF.

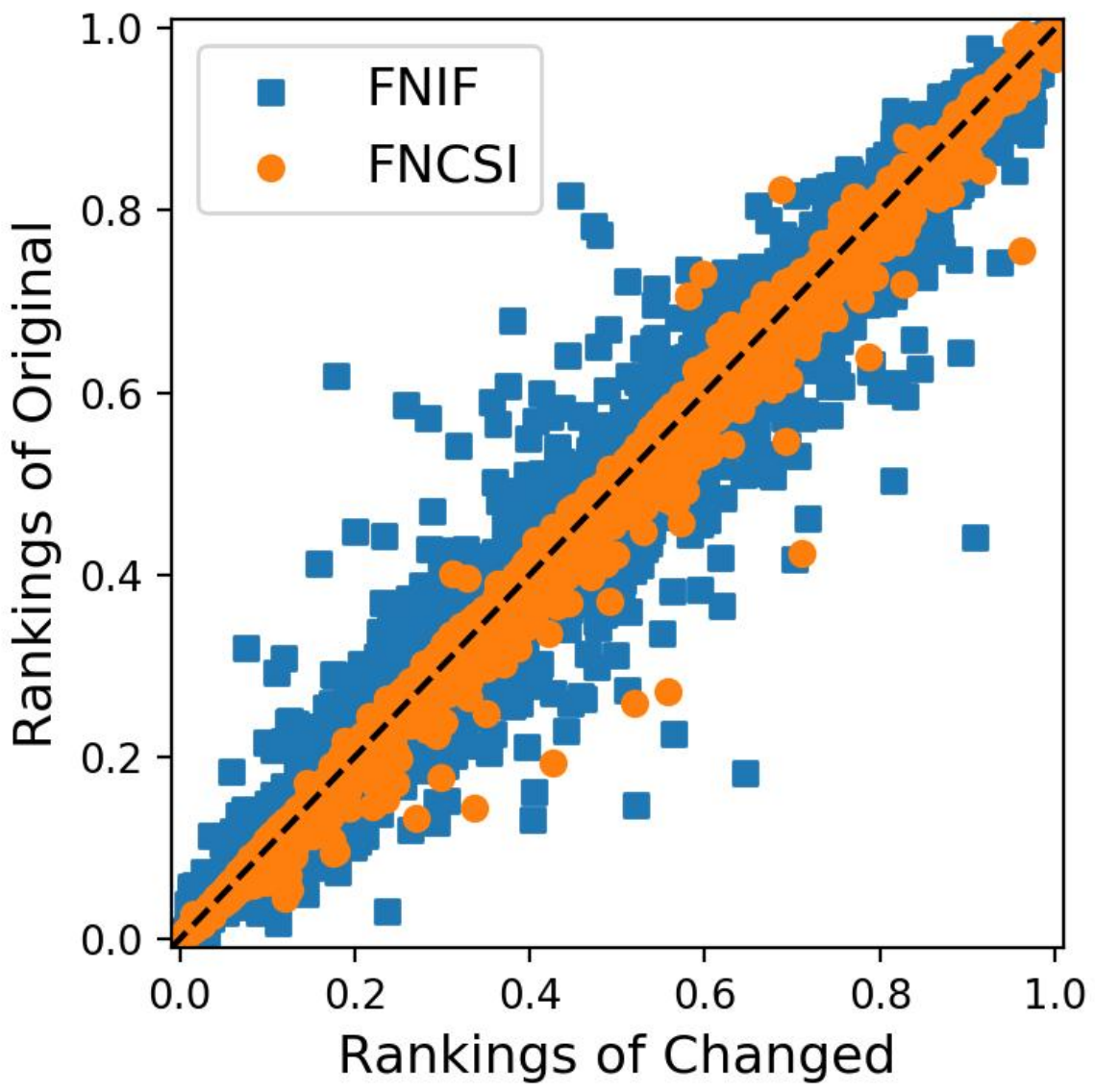

**Figure 6: Robustness against document type for FNCSI and FNIF.**

# 5. Conclusion and Discussion

In this paper, we first reviewed the history of the CAS Journal Ranking. We also listed the main limitations in earlier editions of CAS Journal Ranking, e.g., the old factor (the three-year average JIF) is not robust enough against occasional highly-cited papers; the old discipline categories are not fine-grained enough to achieve a more sophisticated field normalization. To improve these deficiencies, in 2019, two critical improvements were adopted in the upgraded version: the CWTS paper-level classification system is utilized to replace the JCR subject category; FNCSI was introduced to replace JIF. Furthermore, a comparison of FNCSI and FNIF showed that FNCSI successfully addressed the issues mentioned above to a large extent.

The measures we applied in the upgraded version of CAS Journal Ranking offer a novel idea to evaluate journal impact. In this study, the reported results demonstrate that our measures are effective overall. We also received quite some positive feedback (received through internal communication via e-mails, telephone, social media, and interviews) from Chinese researchers. For example, some researchers stated that in the upgraded version, the ranking of journals is more similar to their subjective feelings, and some "good" journals from basic research fields with low potential citations, are revealed in the upgraded version.

### 5.1 Future Research of CAS Journal Ranking

As far as future research is concerned, we propose the following:

- Eliminate the influence on the differentiated distribution of the FNCSI score among disciplines.

FNCSI is a normalized indicator by the discipline of CWTS paper-level classification. Theoretically, the scores of FNCSI can be compared across different disciplines. However, based on the observation of the FNCSI score among disciplines, the distribution of the FNCSI score is differentiated among disciplines (see Appendix E). For example, in this situation, when the journals are divided into different disciplines, their rankings will be different. However, the basis and criteria for assigning journals to different disciplines are imprecise. Therefore, next, we will investigate how to eliminate the influence of this problem.

- Determine the optimal number of tiers

As mentioned in the introduction, for both the earlier and upgraded version, we adopted four tiers. The initial motivation for using four tiers is empirical: the more the number of tiers, the lower the distinction among different tiers and vice versa. Further, we found that four tiers lead to a better distinction in some disciplines. After adopting FNCSI, we started to rethink whether using four tiers is the best option. Why isn’t it three or five tiers? Particularly, for the FNCSI score, optimal solutions for different disciplines may be different. The essential requirement for comparing disciplines is that they all have the same number of tiers. How to balance tradeoffs and find the optimal solution?

- Explore the influence of the paper-level classification system and FNCSI, respectively

There are two main measures adopted in the upgraded CAS Journal Ranking. A fine-grained paper-level classification system and the new field-normalized indicator FNCSI were utilized on the upgraded CAS Journal Ranking. Theoretically, both the classification system and the indicator will influence the ranking results. It is valuable to detect the respective consequence of the two factors and to compare which one has the most significant effect. In further studies, we will explore this point and investigate the other properties of CAS Journal Ranking, e.g., the robustness towards covidization (Zhang, 2022; Liu, 2023a; Liu, 2023b).

In addition, in this paper, although we didn't focus on discussing the issue of research performance evaluation and how to use journal ranking properly, we are on the way to correcting the improper use of CAS Journal Rankings. One measure that has been taken is extending journal rankings to journal profiles based on CWTS paper-level classification and FNCSI, which will provide comprehensive information rather than only metrics.

**5.2 Discussion on the use of CAS Journal Ranking**

CAS Journal Ranking, a fully bibliometric ranking system, should be used following the common principles of bibliometrics, e.g., the compatibility of objective data and peer review across different levels of granularity. Figure 7 illustrates the consensus among scholars in the realm of research evaluation. When evaluating meso- and macro-level entities like research institutions and disciplinary domains, bibliometric indicators, like citation impact-based journal evaluation, provide more reliable information. Conversely, when evaluating micro-level entities such as individual researchers, peer review should play a leading role and bibliometrics should play a supporting role.

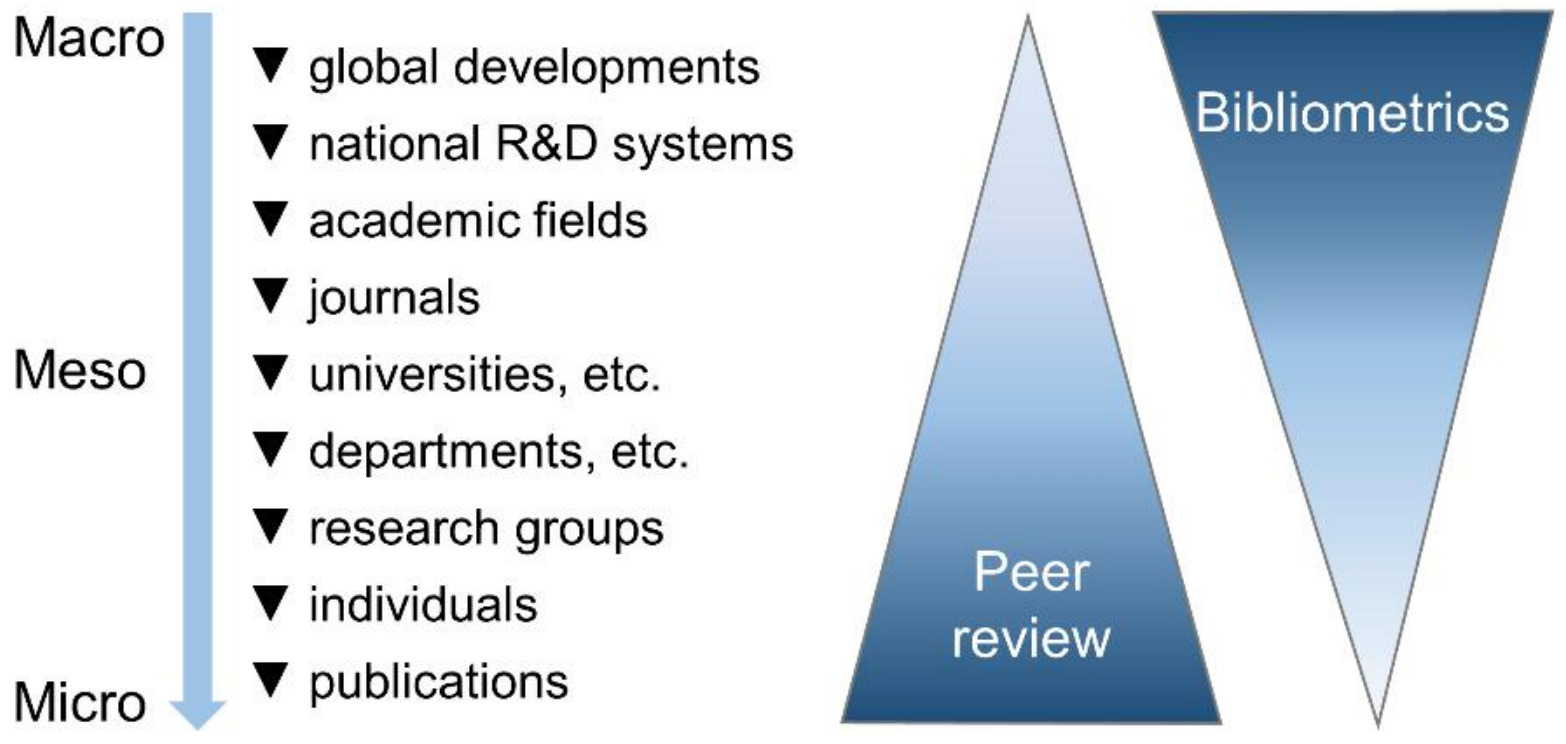

Figure 7. Suitability of bibliometrics and peer review at different levels (based on Hinze, S (2014)). Accordingly, the use of the CAS Journal Ranking can be aggregated into different levels.

- **At the Macro- or Meso-level**

  i. Comparing the research performance of countries or institutions based on the statistics of rankings or tiers of the journals published, like the role played by the *Nature Index*. A better journal ranking system will provide a more accurate macro- or meso-level estimation of performance.

  ii. Facilitating librarians to select which journals to subscribe to. A better journal ranking system may help librarians to better allocate the resources for subscriptions.

iii. Facilitating researchers to select journals for paper reading or manuscript submission. Researchers can start their literature survey from papers published in high-ranking journals and trace the citation flow then. This usage is especially for junior researchers or graduates who are not fully familiar with all the journals in their areas.

- **At the Micro-level**

i. Evaluating researchers (e.g., for promotion, rewards, grant funding) based on the rankings of the journals they used for their publications, e.g., using the number or fraction of highly ranking journals, using the total score transformed from the ranking of journals (Quan et al., 2017).

ii. Evaluating the quality or impact of papers based on the rankings of the journals they published in, e.g., selecting the best paper in an area.

These micro-level direct utilizations of the CAS Journal Ranking are not recommended. In recent years, China's MOE and MOST have released policies to implement reforms to encourage more qualitative evaluation of research. Conducting high-quality peer review requires a set of supporting measures, such as establishing clear guidelines and accountability on peer review, providing education and training on high-quality peer review, and educating on the proper use of quantitative metrics to become bibliometric-wise. In addition, the path to responsible evaluation requires efforts by all stakeholders.

# Acknowledgements

The authors thank the generous help from CWTS (The Center for Science and Technology Studies of Leiden University, Netherlands) and Dr. Nees Jan van Eck for providing the paper-level classification data, and thank Ms. M. Zhu and Dr. Ronald Rousseau for valuable discussion. The authors would also like to thank the editor and anonymous reviewers for their helpful and constructive comments on the manuscript.

# Author contribution

Sichao Tong: Formal analysis, Investigation, Methodology, Visualization, Writing – original draft, Writing – review & editing

Fuyou Chen: Data curation, Writing – review & editing

Liying Yang: Conceptualization, Methodology, Writing – review & editing, Supervision

Zhesi Shen: Conceptualization, Data curation, Formal analysis, Investigation, Methodology, Visualization, Writing – original draft, Writing – review & editing

## Competing interests

The authors declare that they have no conflict of interest.

## Funding information

This study was partly funded by The National Social Science Foundation of China "Research on Semantic Evaluation System of Scientific Literature Driven by Big Data" (Grant No. 21&ZD329).

## Data availability

All journals' FNCSI results for 2019 version are available at 10.57760/sciencedb.08419.

## Appendix

## Appendix A. Examples of FNCSI formulation represented in an analogy way

Figures A1 and A2 show FNCSI in a very intuitive way. The boxes represent journals, the fruits in the boxes represent papers, their size indicates the number of citations of the papers, and their type indicates the research topic and document type of the papers. The FNCSI of box A is the probability that a random fruit from box A is larger than one from the other boxes with the same fruit type.

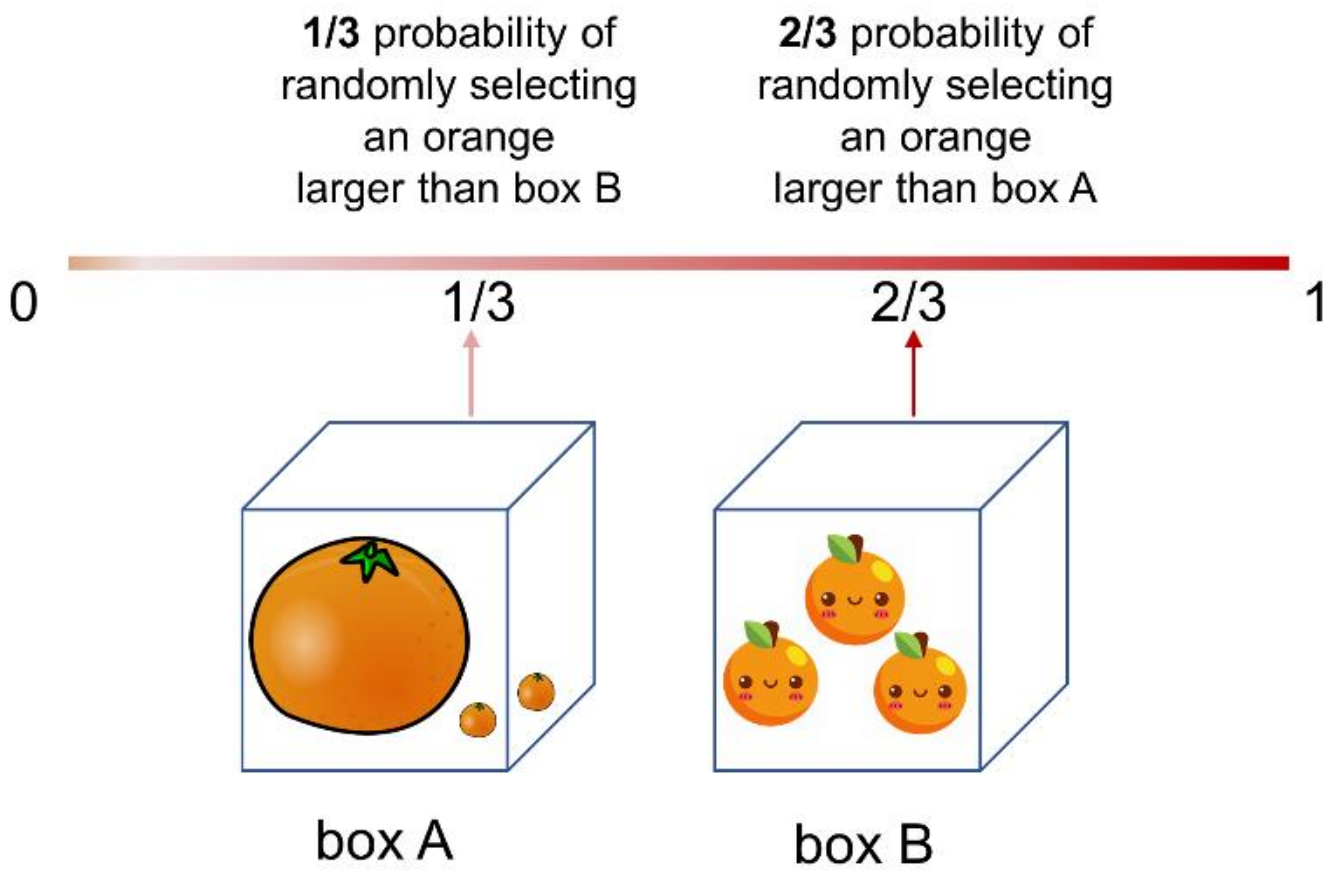

(a)

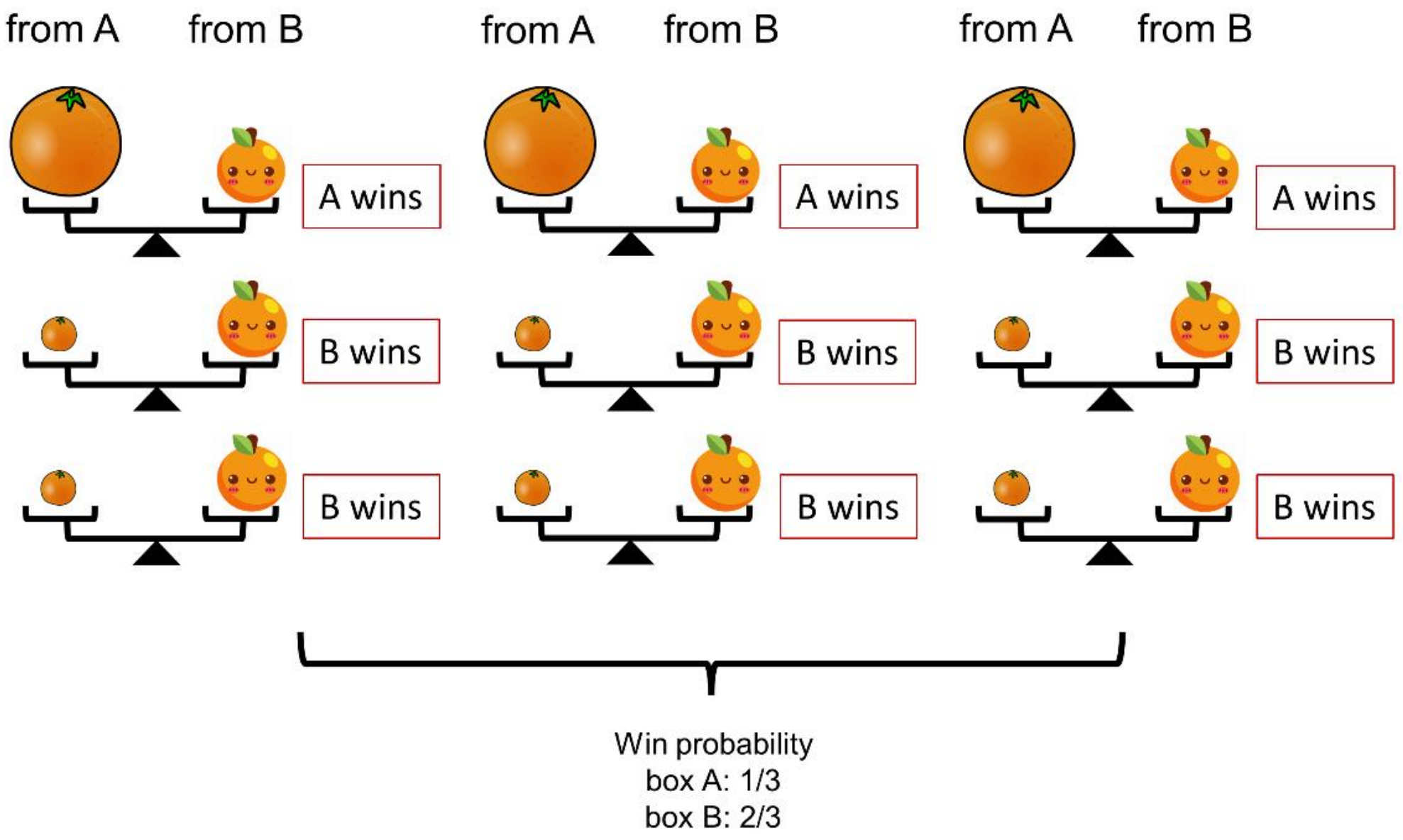

(b)

Figure A1. A metaphorical explanation of FNCSI as fruit size comparison of two boxes. a) Win probability of box A and B; b) Pairwise comparison to obtain the probability.

Next, we use different fruit types to illustrate comparison when taking research topics and document types into account, as shown in Fig. A2.

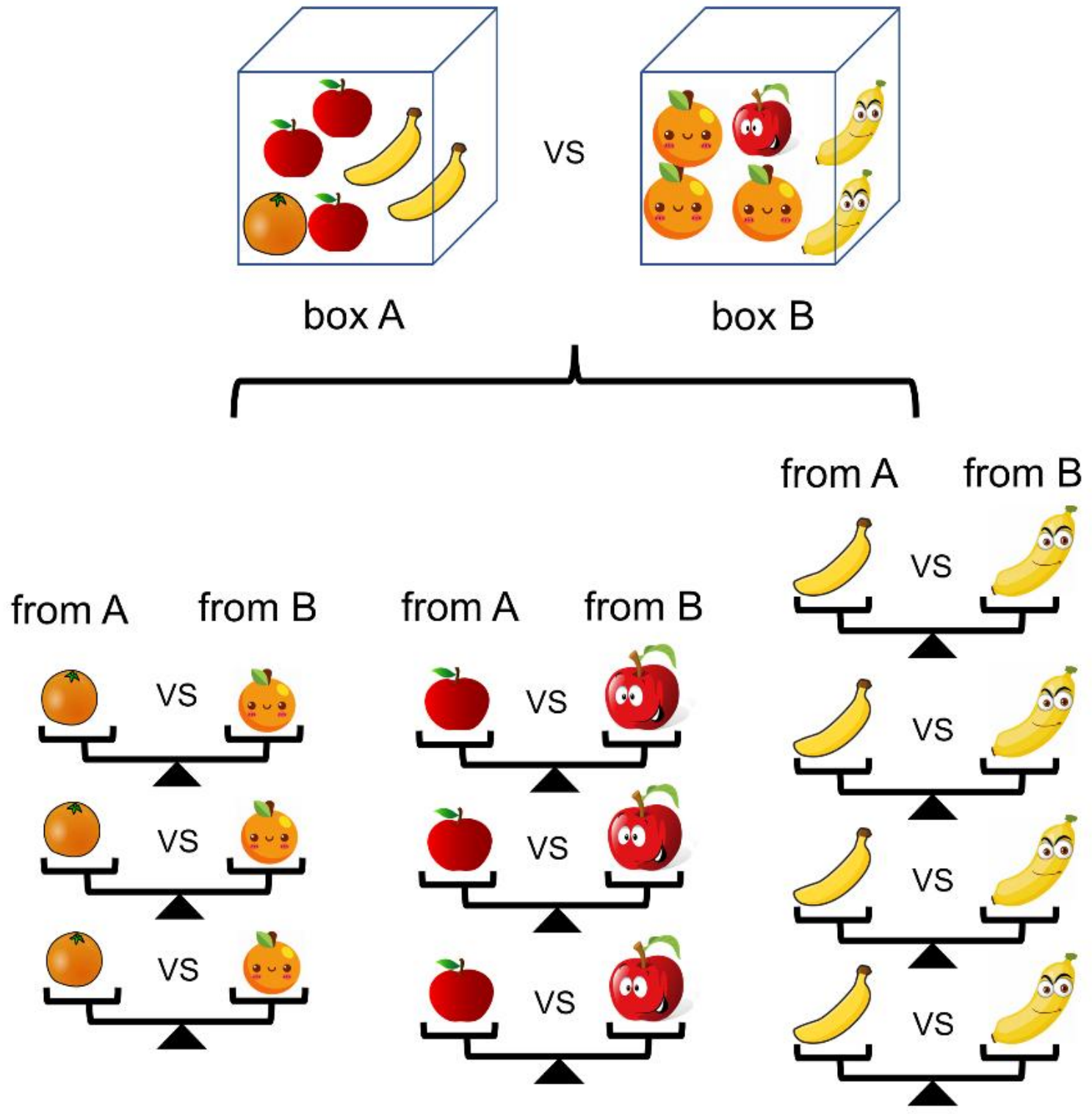

Figure A2. Comparing the size of fruits in two boxes taking fruit types into account.

# Appendix B. Why using FNCSI and its connections to other indicators

The reason why we use FNCSI (or its original version CSI) can be traced back to the previous work we published (Shen, Z., Yang, L., & Wu, J, 2018; Shen, Z., Yang, L., Di, Z., & Wu, J., 2019). In these two studies, we focus on the problem of comparing journals' citations using their mean citation (such as done in the calculation of the JIF), i.e., the relationship between $\overline{X} \succ \overline{Y}$ and $P(X \succ Y)$. In fact, $\mathrm{CSI} = P(X \succ Y)$. So following this research line, here we use FNCSI in the new CAS Journal Ranking. In the following tables, we compared FNCSI with other ranking-based indicators (PPtop10% and Average Percentile are used here). Here we define ranking-based indicators as those indicators whose value changes directly depend on the changes in the ranking of the elements they are calculated from, rather than the changes in the element values themselves. We can see a high correlation among FNCSI, PPtop10%, and Average Percentile, especially between FNCSI and Avg Percentile. PPtop10% mainly focuses on the top cited papers, while FNCSI and Average Percentile mainly focus on the central tendency of the citations of a journal's publication. In fact, we can formulate a theoretical relation between FNCSI and Avg Percentile as follows,

$$\text{FNCSI}_A = 1 - \frac{\text{AvgPer}_A * N_{A+O} + 0.5 N_A}{N_O}$$

where $\text{FNCSI}_A$ is the FNCSI of journal A, $\text{AvgPer}_A$ is the Average Percentile of journal $A$, $N_A$ is the number of publications of journal $A$, $N_O$ is the number of publications of other journals, $N_{A+O}$ is the number of publications of all journals. More detailed connections between FNCSI and other journal indicators can be found in the study by Shen and collaborators (2023).

**Table B1: Spearman Correlation of journal indicators.**

| | JIF | FNIF | FNCSI | PPtop10% | Avg Percentile |
|---|---|---|---|---|---|
| **JIF** | 1 | 0.83 | 0.82 | 0.75 | 0.93 |
| **FNIF** | | 1 | 0.98 | 0.95 | 0.95 |
| **FNCSI** | | | 1 | 0.91 | 0.96 |
| **PPtop10%** | | | | 1 | 0.87 |
| **Avg Percentile** | | | | | 1 |

**Table B2: Pearson Correlation of journal indicators.**

| | JIF | FNIF | FNCSI | PPtop10% | Avg Percentile |
|---|---|---|---|---|---|
| **JIF** | 1 | 0.82 | 0.57 | 0.67 | 0.60 |
| **FNIF** | | 1 | 0.81 | 0.91 | 0.80 |
| **FNCSI** | | | 1 | 0.84 | 0.97 |
| **PPtop10%** | | | | 1 | 0.82 |
| **Avg Percentile** | | | | | 1 |

# Appendix C. Top 20 ranked research journals

In Table C1, we list the top 20 ranked research journals based on FNCSI and FNIF, respectively. Compared with the journals selected according to FNCSI, the top four journals via FNIF are all medical-related.

**Table C1: Top 20 journals based on FNCSI and FNIF respectively.**

| Journal | FNCSI | journal | FNIF |
|---|---|---|---|
| LANCET | 1 | CA-CANCER J CLIN | 1 |
| NATURE | 2 | NEW ENGL J MED | 2 |
| JAMA-J AM MED ASSOC | 3 | LANCET | 3 |
| SCIENCE | 4 | JAMA-J AM MED ASSOC | 4 |
| CELL | 5 | NATURE | 5 |
| WORLD PSYCHIATRY | 6 | PSYCHOL SCI PUBL INT | 6 |
| LANCET NEUROL | 7 | Q J ECON | 7 |
| NAT PHOTONICS | 8 | WORLD PSYCHIATRY | 8 |
| NAT GENET | 9 | SCIENCE | 9 |
| NAT MED | 10 | LANCET ONCOL | 10 |
| NAT MATER | 11 | LANCET NEUROL | 11 |
| LANCET ONCOL | 12 | NAT MATER | 12 |
| CANCER CELL | 13 | NAT GENET | 13 |
| NAT CHEM | 14 | PSYCHOL BULL | 14 |
| NAT NEUROSCI | 15 | CELL | 15 |
| CELL METAB | 16 | NAT ENERGY | 16 |
| LANCET RESP MED | 17 | NAT PHOTONICS | 17 |
| NAT IMMUNOL | 18 | CIRCULATION | 18 |
| LANCET DIABETES ENDO | 19 | FUNGAL DIVERS | 19 |
| NAT NANOTECHNOL | 20 | LANCET INFECT DIS | 20 |

# Appendix D. Additional results on robust comparison between FNCSI and FNIF

In this section, we present some additional results on the robustness of the proposed journal indicators. In Fig.4(b), we have illustrated the relative change of rankings based on FNCSI and FNIF, here we demonstrate some further results. In Fig.D1 we compare the 1st quartile and 3rd quartile rankings obtained from the 100 simulations for each journal. The x-axis is the 1st quartile and the y-axis is the 3rd quartile. For a robust indicator, the 1st quartile and 3rd quartile should be pretty close, thus the dot will be located close to the diagonal line. We can see that for both FNCSI and FNIF, the dots are mainly located along the diagonal line implying that the rankings of most journals are stable. When comparing the orange dots (FNCSI) and blue squares (FNIF), we can see that the area over which orange dots are spread is smaller than that of the blue squares, indicating that rankings based on FNCSI are more stable than rankings based on FNIF when dealing with some special journals.

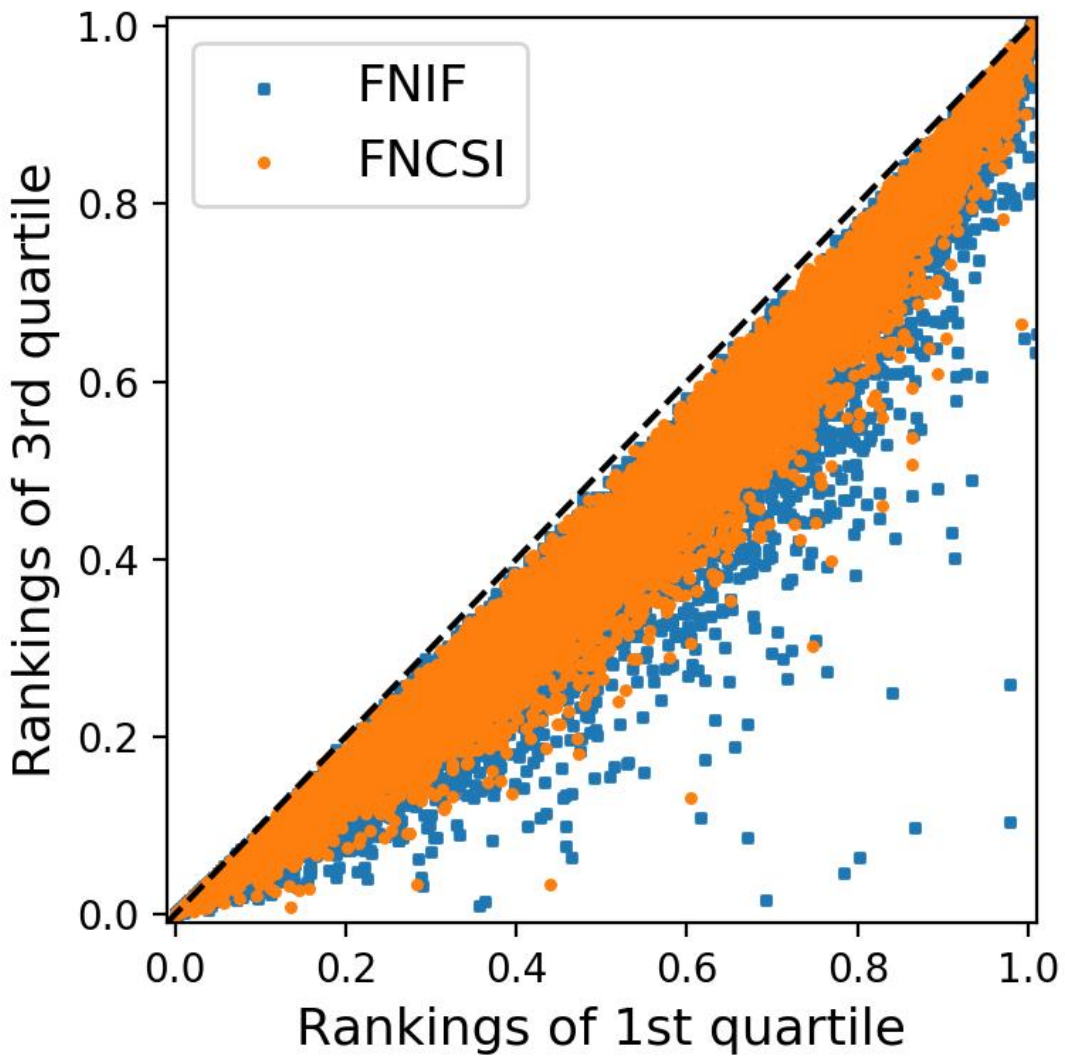

**Figure D1: Change of rankings based on FNCSI and FNIF.**

Journal indicators should also be stable over time as a journal's reputation and quality will not change dramatically in a short time. In Fig.D2 we present, as an example, the evolution of rankings based on JIF, FNIF, and FNCSI for the journal *J Math Sociol*. We can see the rankings of JIF and FNIF show a big jump in the year 2018 compared with their rankings in the previous and following years. However, the ranking of FNCSI only increases slightly.

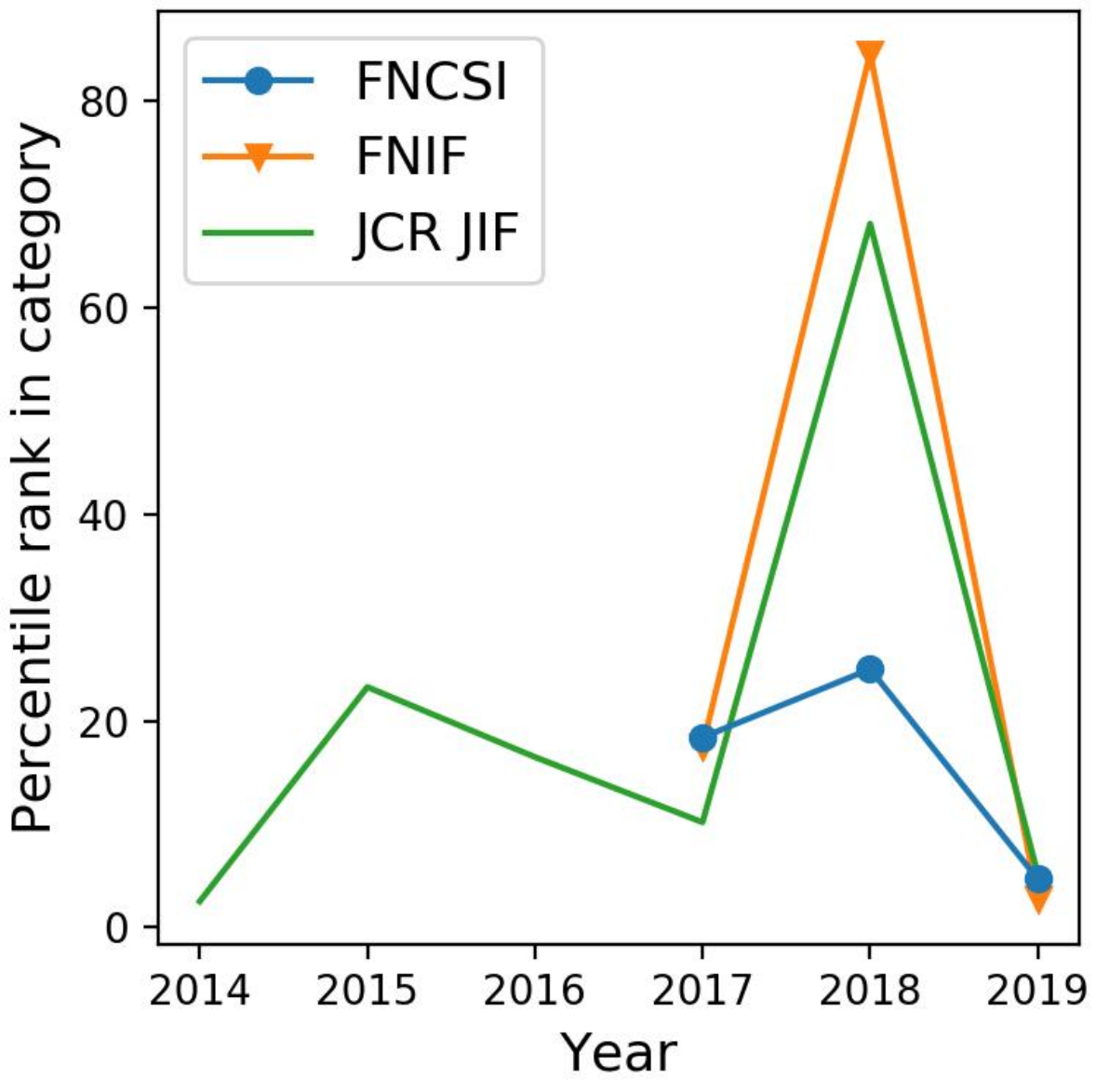

**Figure D2: Evolution of percentile rank for *J Math Sociol* based on different indicators. The percentile ranking is calculated within the Mathematics, Interdisciplinary Applications category.**

# Appendix E. Distribution of journals' FNCSI in different disciplines

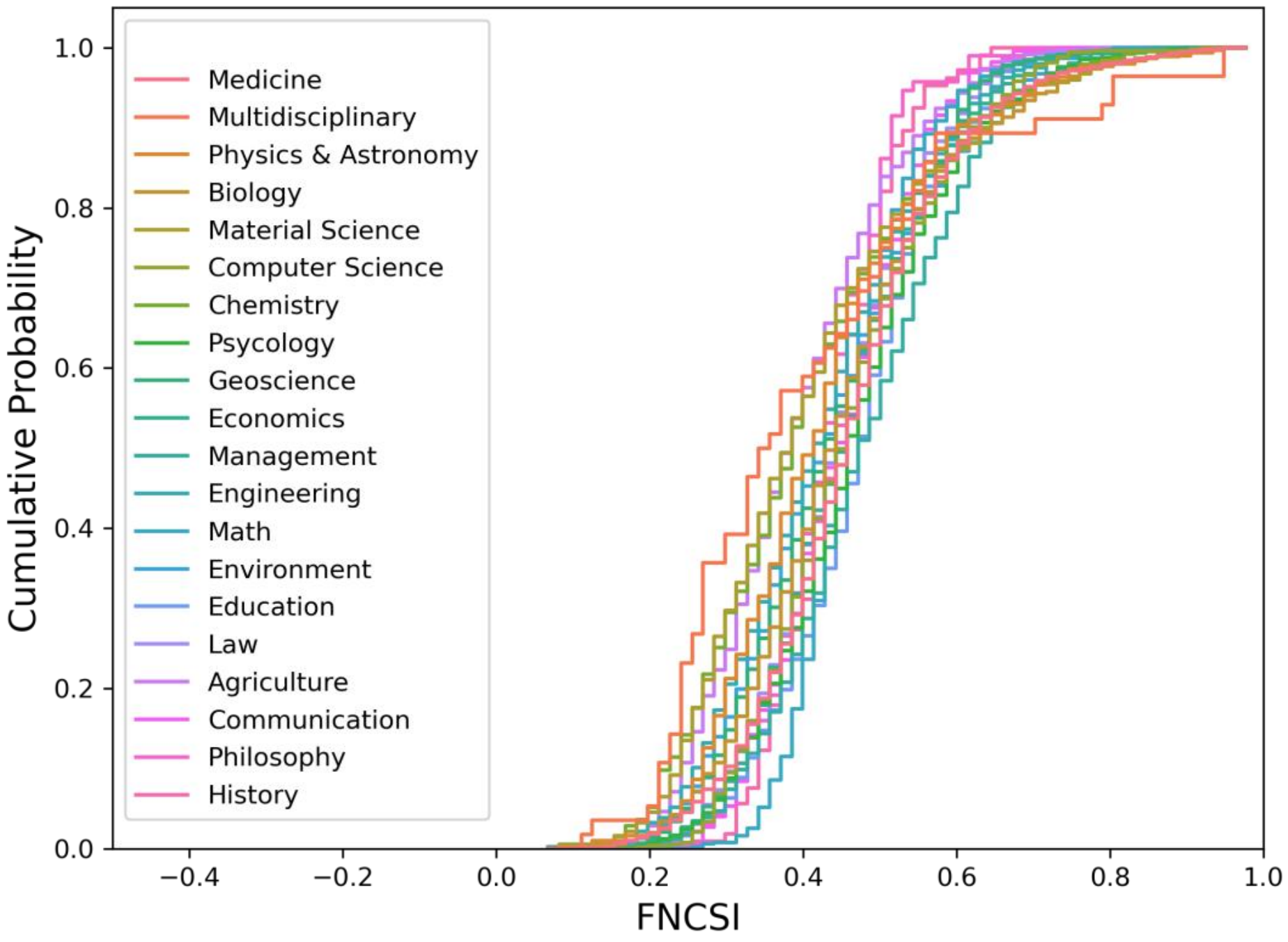

**Figure E1: Cumulative probability distribution of journals' FNCSI of different disciplines.**